%% file: paper.tex
\PassOptionsToPackage{dvipsnames}{xcolor}
\documentclass[10pt,conference]{IEEEtran}
\IEEEoverridecommandlockouts
\usepackage{cite}
\usepackage{amsmath,amssymb,amsfonts}
\usepackage{algorithmic}
\usepackage{graphicx}
\usepackage{textcomp}
\usepackage{xcolor}
\usepackage{balance}
\usepackage{booktabs}
\usepackage[all]{nowidow}

\usepackage[hidelinks]{hyperref}
\usepackage[capitalise]{cleveref}

\newcommand{\gf}{\textit{GeoFaaS}}

\begin{document}

\title{GeoFaaS: An Edge-to-Cloud FaaS Platform \\
    \thanks{Funded by the Bundesministerium f{\"u}r Bildung und Forschung (BMBF, German Federal Ministry of Education and Research) -- 16KISK183.}
}

\author{\IEEEauthorblockN{Mohammadreza Malekabbasi, Tobias Pfandzelter, Trever Schirmer, David Bermbach}
    \IEEEauthorblockA{\textit{Technische Universit\"at Berlin \& Einstein Center Digital Future}\\
        \textit{Scalable Software Systems Research Group} \\
        \{mm,tp,ts,db\}@3s.tu-berlin.de}
}

\maketitle

\begin{abstract}
    The massive growth of mobile and IoT devices demands geographically distributed computing systems for optimal performance, privacy, and scalability.
    However, existing edge-to-cloud serverless platforms lack location awareness, resulting in inefficient network usage and increased latency.
    
		In this paper, we propose \gf{}, a novel edge-to-cloud Function-as-a-Service (FaaS) platform that leverages real-time client location information for transparent request execution on the nearest available FaaS node.
		If needed, \gf{} transparently offloads requests to the cloud when edge resources are overloaded, thus, ensuring consistent execution without user intervention.
    \gf{} has a modular and decentralized architecture: building on the single-node FaaS system tinyFaaS, GeoFaaS works as a stand-alone edge-to-cloud FaaS platform but can also integrate and act as a routing layer for existing FaaS services, e.g., in the cloud.
        To evaluate our approach, we implemented an open-source proof-of-concept prototype and studied performance and fault-tolerance behavior in experiments.
\end{abstract}


\input{sections/1_introduction}
\input{sections/2_related_work.tex}
\input{sections/3_architecture}

\input{sections/4_evaluation.tex}
\input{sections/5_conclusion.tex}

\balance

\bibliographystyle{IEEEtran}
\bibliography{bibliography-ieee}

\end{document}

%% file: sections/1_introduction.tex
\section{Introduction}
\label{sec:introduction}
The field of edge and cloud computing holds significant promise for various applications.
However, there is a tangible lack of applicability in practice since building and managing an edge-to-cloud system can be complex, hindering widespread adoption~\cite{mohan2020pruning,paper_bermbach2017_fog_vision}.
Due to its abstractions, Function-as-a-Service (FaaS) promises to be a suitable approach for an edge-to-cloud application runtime environment~\cite{paper_pfandzelter2020_tinyfaas,george2020nanolambda,russo2023serverledge,oliveira2023function, paper_bermbach2021_auctionwhisk,raith2023serverless, aslanpour2021serverless,paper_bermbach2021_cloud_engineering}:
Developers only need to write individual, by now even stateful~\cite{pfandzelter2023enoki}, code functions while the rest of the application lifecycle is handled by the FaaS provider.

Despite these obvious benefits, current state-of-the-art edge-to-cloud FaaS platforms, e.g.,~\cite{paper_pfandzelter2020_tinyfaas,george2020nanolambda,russo2023serverledge,baresi2019towards} lack a key feature: geographical context awareness~\cite{oliveira2023function}.
Considering the geographical location of potentially mobile serverless clients when routing their requests to the nearest serverless node can reduce network traffic and facilitate load balancing~\cite{aslanpour2021serverless}.
Additionally, in a changing network, physical distance serves as a simple yet effective approximation for network latency~\cite{paper_hasenburg2020_disgb, ng2002predicting}.

This paper addresses the limitation of inherent geographical context awareness in FaaS platforms.
We propose \gf{} (Geo-aware Function-as-a-Service), a geographically distributed FaaS platform operating across the heterogeneous edge-to-cloud continuum.
By leveraging client location information, GeoFaaS transparently routes function requests to the nearest available FaaS resource, minimizing network latency for improved performance.
Additionally, \gf{} supports transparent offloading by means of fallback processing in the cloud to handle edge failures resulting from, e.g., overload situations. 
Building on the single-node FaaS system tinyFaaS~\cite{paper_pfandzelter2020_tinyfaas}, GeoFaaS works as a stand-alone edge-to-cloud FaaS platform.
Nevertheless, GeoFaaS can use its modular architecture to integrate existing FaaS services, e.g., in the cloud, and also act as a routing layer only.

In summary, we make the following contributions:
\begin{itemize}
    \item We propose \gf{}, an edge-to-cloud FaaS platform (\cref{sec:architecture}).
    \item We implement a proof-of-concept prototype and the corresponding evaluation setup, both available as open-source software\footnote{\url{https://github.com/OpenFogStack/GeoFaaS}} (\cref{subsec:implementation}).
    \item We demonstrate performance and fault-tolerance characteristics of \gf{} in three experiment scenarios (\cref{subsec:experiments}). 
\end{itemize}

%% file: sections/2_related_work.tex
\section{Background and Related Work}
\label{sec:related}

FaaS systems can facilitate edge computing by being lightweight enough for resource-constrained environments and flexible enough for heterogeneous hardware.
Systems like tinyFaaS~\cite{paper_pfandzelter2020_tinyfaas}, Sledge~\cite{gadepalli2020sledge}, and CSPOT~\cite{wolski2019cspot} exemplify this approach.
However, these systems only consider running on a single node, which is inadequate for most real-world applications.
Therefore, interconnecting FaaS nodes offers a solution to meet the demands of practical use cases.

There has been a growing interest in edge-to-cloud FaaS in recent works~\cite{oliveira2023function}.
Baresi et al.~\cite{baresi2019towards} discuss a hierarchical FaaS architecture that offloads from edge/fog servers to the cloud, using a centralized orchestrator.
Russo et al.~\cite{russo2023serverledge} present Serverledge, a decentralized approach where edge servers within the same pre-defined group, determined by their proximity, collaborate for horizontal offloading.
Although Serverledge aligns with our work in decentralization, it lacks the capability to guarantee client connections to the nearest server or nearest edge zone.
This presents a hurdle, particularly for mobile clients whose physical location and proximity to edge nodes changes frequently.
Further, the Serverledge cloud offloading necessitates setting the explicit address of cloud for each edge node.
Cicconetti et al.~\cite{cicconetti2020decentralized} present a decentralized FaaS framework for the edge, which routes the requests depending on the connectivity of the routing brokers to the FaaS nodes and the past execution history.
Ciavotta et al.~\cite{ciavotta2021dfaas} use a peer-to-peer overlay network of routing brokers to make a distributed FaaS platform for the edge.
While numerous other studies exist~\cite{hetzel2021muactor, smith2022fado, lordan2021colony}, to our knowledge, none consider clients' locations in their request routing.
This opens up promising possibilities for performance and client experience improvements.

Hasenburg et al.~\cite{paper_hasenburg2020_geobroker} present \textit{GeoBroker}, a publish/subscribe system that leverages the geographical location of clients and designated areas called \textit{geofences}.
In their system, publishers must specify a target geofence to reach relevant subscribers within that specific area.
This targeted message delivery is achieved through a geo-filtering process called \textit{event geo-check}, ensuring subscribers only receive publications relevant to their location.
Similarly, subscribers use the same mechanism to limit their subscription to a targeted area of publishers.
In another study, Hasenburg et al.~\cite{paper_hasenburg2020_disgb} use Rendezvous Points (RP) for dissemination of events and subscriptions in distributed publish/subscribe systems.
They introduce \textit{broker areas}, meaning that each GeoBroker serves only its own distinct, non-overlapping assigned area.
These systems can offer geo-context awareness; however, to our knowledge, they have not yet been integrated with serverless and FaaS systems.
While client-side location measurement equipment is necessary, the client needs no interaction with the system for location determination, and most clients already possess this equipment for other purposes.
For sake of brevity, we call a single broker node ``GeoBroker''~\cite{paper_hasenburg2020_geobroker}, and distributed GeoBroker ``DisGB''~\cite{paper_hasenburg2020_disgb}.

%% file: sections/3_architecture.tex
\section{GeoFaaS Architecture}
\label{sec:architecture}
\begin{figure}
    \centering
    \includegraphics[width=\linewidth]{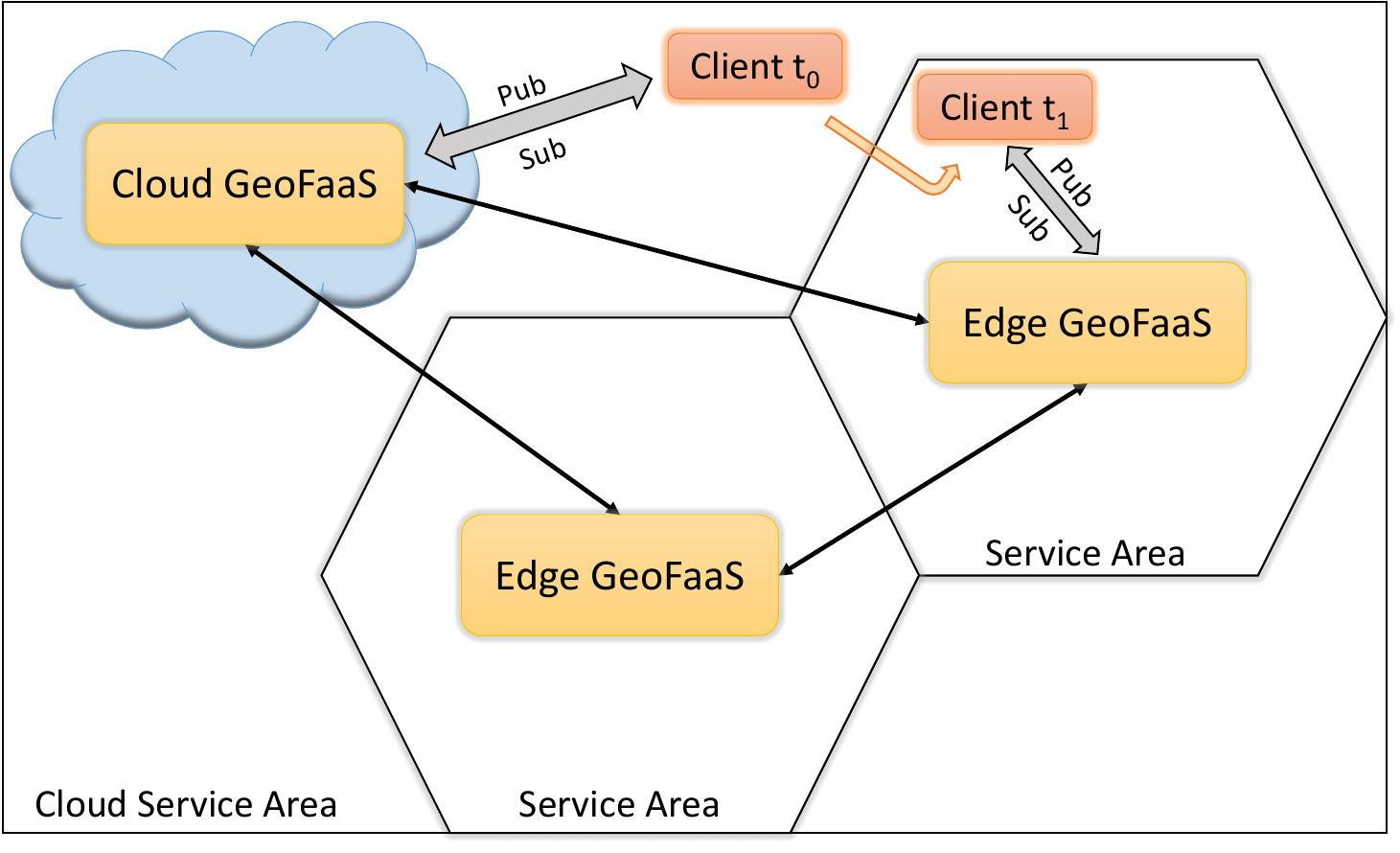}
    \caption{The \gf{} system with two \gf{} nodes deployed on edge and one node deployed on the cloud. The Client at time $t_0$ is outside of edge service areas, so it communicates with cloud. When client moves to a new service area at $t_1$, the cloud node will hand off the client to a new node for further communication by its last responsible node. %
    }
    \label{img:systemdesign}
\end{figure}

\begin{figure}
    \centering
    \includegraphics[width=\linewidth]{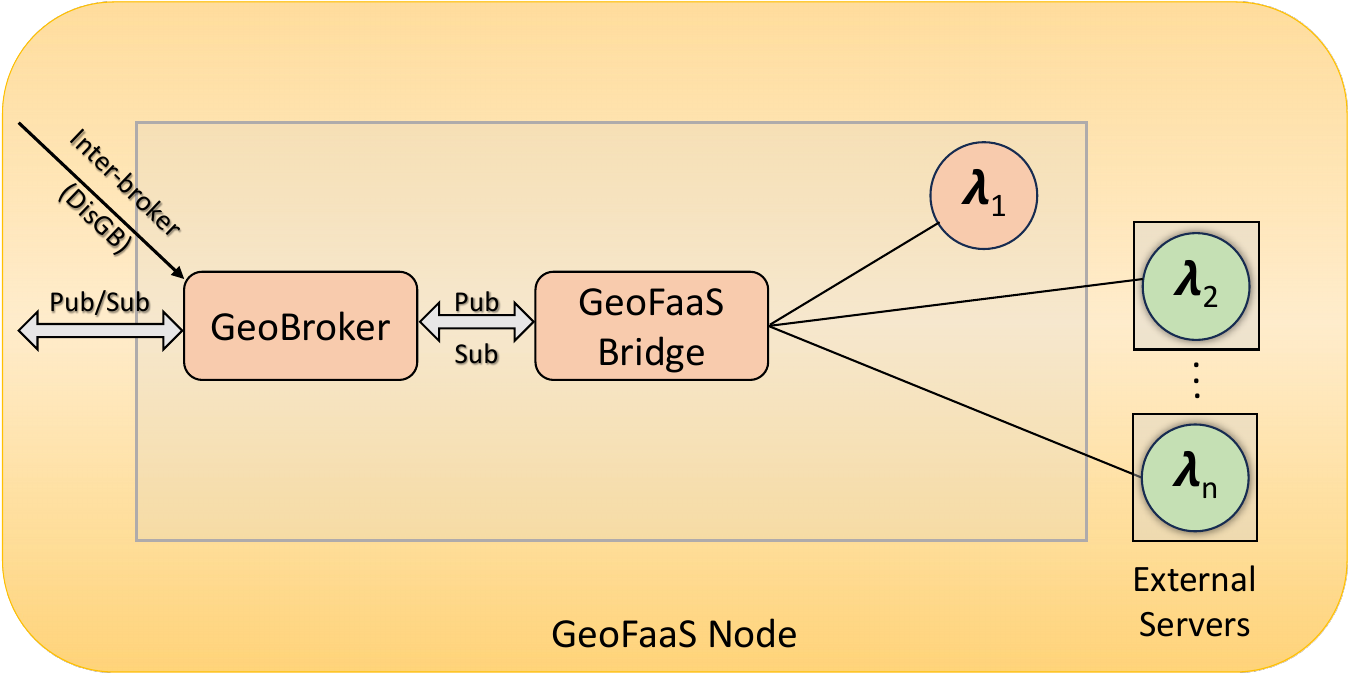}
    \caption{The \gf{} node architecture comprises three key components: GeoBroker, which enables location-aware pub/sub messaging, a local FaaS server and/or FaaS services running on external servers within the same data center, and the \gf{} Bridge, a middleware between the other components.}
    \label{img:nodedesign}
\end{figure}

Inspired by FBase and FReD~\cite{pfandzelter2023fred,techreport_hasenburg2019_fbase,paper_hasenburg_towards_fbase}, GeoFaaS utilizes a node concept, grouping one or more servers at a specific location within the edge-cloud continuum -- see \cref{img:systemdesign}.
While requiring at least one node, GeoFaaS typically leverages numerous nodes distributed across edge and cloud.
Edge nodes manage non-overlapping \textit{service areas}, defining their responsibility for clients located within that specific geographical region, a concept introduced by DisGB~\cite{paper_hasenburg2020_disgb}.
The cloud acts as a backup mechanism, dynamically handling edge node failures due to FaaS resource limitations or internal crashes.

Each \gf{} node has three key components: an instance of GeoBroker (\ref{subsec:architecture:geobroker}) as message broker, FaaS servers (\ref{subsec:architecture:faas}) for function execution, and the GeoFaaS Bridge (\ref{subsec:architecture:bridge}) for connecting the aforementioned elements (See also \cref{img:nodedesign}).
Beyond these, there is also a client-side library, the \gf{} clients (\ref{subsec:architecture:clients}), which handles interaction between application clients and GeoFaaS platform.

\subsection{Distributed Geo-Context Aware Message Broker}
\label{subsec:architecture:geobroker}

Within a \gf{} node, a GeoBroker instance is co-located with a \gf{} Bridge instance  (see \cref{img:nodedesign}), either on the same physical machine or within the same data center network.
The GeoBroker instances running on all the \gf{} nodes form the distributed GeoBroker, DisGB.
Leveraging a publish/subscribe paradigm and topics, DisGB facilitates communication between \gf{} clients, the \gf{} Bridge, and intra-node communication, eliminating the need for direct IP address usage.
Notably, it establishes a geographically aware distributed publish/subscribe infrastructure, to disseminate information based on the location of publishers and subscribers.
For DisGB to function, all GeoBroker instances must have the same version of node registry.
We assume this registry to be relatively static in real use cases, and updates can be fetched from a global registry if necessary.
Furthermore, we set GeoBroker's RPs to be near the subscribers.

In order for a client to initiate communication with the system, it has the freedom to target any GeoBroker instance (no matter where, possibly cached from a previous request or fetched from DNS) to publish its first request.
Then DisGB acts as a smart intermediary, directing the request to the geographically closest \gf{} Bridge (subscriber) to the client.
From this point, for any follow-up requests, the client will communicate directly with its responsible GeoBroker instance.
Note that, disjoint edge broker areas are complemented by the cloud's broker area, guaranteeing total coverage.
This means that even when a client transitions beyond the boundaries of its current GeoBroker's area, as depicted in \cref{img:systemdesign}, the current GeoBroker will hand off the client to the new responsible GeoBroker.
This ensures that each client is connected to the geographically most relevant GeoBroker instance for further communication.

It should be noted that, we have deactivated DisGB's event geo-check mechanism for \gf{} Bridge subscription, so that the client does not need to specify the location of the responsible \gf{} node when publishing its request.

\subsection{FaaS Servers}
\label{subsec:architecture:faas}
The modular architecture and extensibility of \gf{} facilitate the concurrent operation of multiple FaaS servers within a single \gf{} node.
For this, GeoFaaS integrates the single-node FaaS system tinyFaaS~\cite{paper_pfandzelter2020_tinyfaas} as the default.
Nevertheless, we explicitly decided on this modular architecture to assert that GeoFaaS can integrate existing FaaS services, e.g., using AWS Lambda for a cloud node instead.

FaaS servers, regardless of their type, expose their available functions through a dedicated endpoint and are called from the \gf{} Bridge which also receives the function responses.
Then for each function running on the FaaS servers, the \gf{} Bridge subscribes to a topic associated with that function name such that clients' requests can reach it.
Similar to AuctionWhisk~\cite{paper_bermbach2021_auctionwhisk,paper_bermbach2020_auctions4function_placement}, we assume that the executables of requested functions have already been deployed independently from GeoFaaS.
Ideally, such function executable would be deployed on all nodes; GeoFaaS, however, will work reliably if the executables have at least been deployed in the cloud.

\subsection{\gf{} Bridge}
\label{subsec:architecture:bridge}
Upon a function request, the \gf{} Bridge selects a FaaS server for function execution based on servers' utilization, availability, and other factors.
If the initially chosen FaaS server fails to respond, the \gf{} Bridge tries the next FaaS server serving the function and if there is no such FaaS server, the \gf{} Bridge will offload the request to the cloud via DisGB.
The \gf{} Bridge employs an autonomous decision-making process, considering various factors, to select a local FaaS server or initiate cloud offloading.
This process remains entirely transparent to both the requesting client and the FaaS servers.
The cloud-deployed \gf{} Bridge subscribes to an ``offloading topic'' for each function, allowing the edge nodes to transparently hand off requests.
This Bridge instance additionally subscribes to a separate ``direct invocation topic'' for each function which can be leveraged by clients to directly invoke a function in the cloud, bypassing the edge if desired.
This topic serves as a fallback option in situations where clients fail calling a function on their responsible edge node.
While the cloud handles the offloaded traffic, directs client retries, and serves requests outside of edge areas, the whole \gf{} system adheres to a decentralized paradigm.

As mentioned before, each \gf{} node has a predefined service area which corresponds to the DisGB broker areas.
The \gf{} Bridge uses this service area as a filter criterion for subscribing to function calls, ensuring only relevant requests are received.
While edge \gf{} nodes mirror their corresponding GeoBroker's broker area as their service area, the service area of the cloud covers the entire world.
Each \gf{} Bridge and its associated GeoBroker instance operate autonomously, making independent decisions within their respective service areas.

\subsection{\gf{} Clients}
\label{subsec:architecture:clients}
Clients use a library as their intermediary for interacting with DisGB that abstracts the underlying implementation details:
to application clients, the library has the same abstraction of synchronous or asynchronous function calls on a non-geodistributed FaaS platform.
Even if the client is within a service area, a request targeting a different \gf{} node is automatically rerouted to the responsible \gf{} node by the system.
The client accordingly gets notified of the \gf{} node to pick up the response from.
Thus, further client requests are directed to the closest \gf{} node to avoid excessive network trips.
This approach provides simplicity and transparency in client communication with the system.

One alternative to centralized cloud monitoring involves client-side retries.
While the cloud can theoretically monitor all function calls and intervene in the event of an edge failure after sending an acknowledgment, this approach raises privacy and bandwidth concerns. Instead, the client can proactively employ a retry mechanism triggered by a predefined timeout.
Moreover, in the event of a \gf{} Bridge failure before a client request is sent, the associated GeoBroker instance automatically redirects the request to the cloud, where the cloud will send the result back to the client through the same GeoBroker instance.
Note that we rely on GeoBroker's heartbeat mechanism to detect \gf{} Bridge's failure.

To summarize, the client makes sure its request is sent successfully, afterwards checks whether \gf{} node has received it and then waits for the function result.
If none was successful, the client tries calling again directly to the cloud \gf{} via the ``direct invocation topic''.

%% file: sections/4_evaluation.tex
\section{Evaluation}
\label{sec:evaluation}

To evaluate GeoFaaS, we implemented an open-source proof-of-concept prototype (\ref{subsec:implementation}) and studied its performance and fault-tolerance behavior in three experiments (\ref{subsec:experiments}).

\begin{table}[h!]
    \centering
    \caption{Topics created by \gf{} for the $f1$ function.}
    \begin{tabular}{@{}llr@{}} \toprule
    \textbf{\#} & \textbf{Topic} & \textbf{Explanation} \\
    \midrule
    1 & \texttt{/f1/call} \label{topic:f1-call} & Client calls function (Bridge subscribes) \\
    2 & \texttt{/f1/ack} \label{topic:f1-ack} & Bridge acknowledges call (Client subscribes) \\
    3 & \texttt{/f1/result} \label{topic:f1-result} & Client subscribes for result (Bridge publishes) \\
    4 & \texttt{/f1/nack} \label{topic:f1-nack} & Edge Bridge offloads call (Cloud subscribes) \\
    5 & \texttt{/f1/call/retry} \label{topic:f1-call-retry} & Client direct cloud call (Cloud subscribes) \\
    \bottomrule
    \end{tabular}
    \label{tab:f1-topics}
\end{table}

\subsection{Proof-of-Concept Implementation}
\label{subsec:implementation}
We implement the \gf{} Bridge and the client library in Kotlin and use the HTTP endpoint of tinyFaaS. %
For each deployed function, we create five topics in DisGB.
In case of an $f1$ function, the specific topics used are detailed in \cref{tab:f1-topics}.
Upon calling a function via \texttt{/call}, the \gf{} client actively listens for both the \texttt{/ack} and \texttt{/result} topics associated with the function, ensuring it receives both an acknowledgment of its request and the function's result.
The cloud is responsible for the requests outside all (edge) service areas, therefore its subscription geofence is the world.
Unlike the cloud, the edge subscribes only to \texttt{/call} topics per function.
As indicated in \cref{tab:f1-topics}, for each function, the cloud subscribes to two additional topics: one for direct client calls in case of unresponsive edges (retry by cloud), and another for edge nodes signaling their inability to serve a request.
In the following experiments, we deactivate ``retry by cloud'', as we expect results to be delivered without needing the \texttt{/call/retry} topic for direct cloud calls.

\subsection{Experiments}
\label{subsec:experiments}
We present a series of experiments to demonstrate GeoFaaS under various conditions.
The next section describes the setup shared by all experiments, subsequent sections explore specific scenarios: \emph{Distance/Latency Change} (\ref{subsubsec:distance}), \emph{High Load} with transparent offloading (\ref{subsubsec:highload}), and system resilience against GeoFaaS Bridge failure (\ref{subsubsec:outage}). %
Each experiment was repeated three times.
As all runs showed similar results, we report the results of a representative run for each scenario.

\subsubsection*{\textbf{Environment Overview}}
\label{subsubsec:environment}
\begin{figure}
    \centering
    \includegraphics[width=\linewidth]{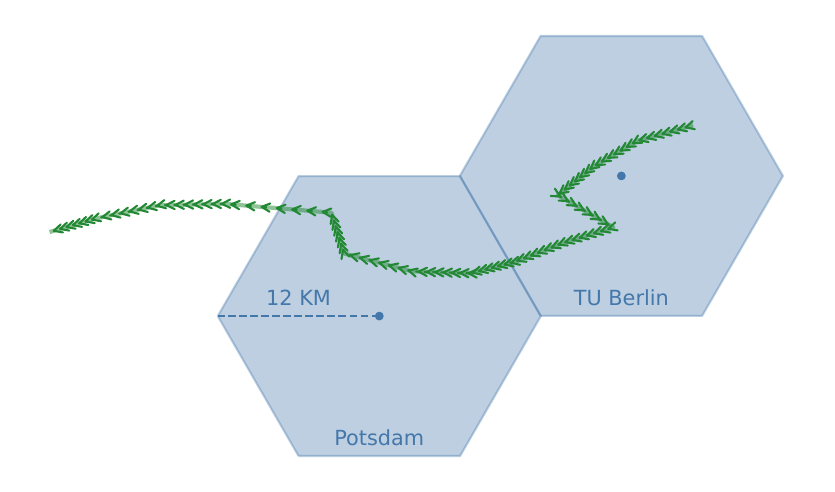}
    \caption{Used service areas for edge brokers in experiments, illustrating client movement as described in section \ref{subsubsec:distance}.}
    \label{img:exp-map}
\end{figure}

For more realistic experiments, the cloud node is deployed on a Google Cloud Platform (GCP) VM with 4 vCPUs and 4 GB of RAM (``e2-highcpu-4''), with almost the same operating system as edge servers.
The cloud node is located in London (``europe-west2-c'').
We measure the average ping latency at 25ms from our local network, covering the path from both the edge network and the client network to the cloud.
Two additional edge nodes are deployed on Raspberry Pi 4 B devices within the local network, each equipped with 4-core ARM processor and 4 GB of RAM.
We define hexagonal service areas with a circumcircle radius of 12 km for edge nodes, as illustrated in \cref{img:exp-map}.
For consistency, clients are also run on an individual Raspberry Pi 4 B unit.
Each mentioned server has their tinyFaaS and GeoBroker instance running locally with the same version of JVM.
To make the setup more realistic, we use \texttt{tc} to inject an artificial 5ms delay into edge-to-edge communication.
Note that all tinyFaaS instances run the ``sieve'' function (prime calculation below 10k), available as a benchmark in tinyFaaS\cite{paper_pfandzelter2020_tinyfaas}.
We observed that the same function runs approximately x2 faster on average on the cloud (3ms vs 6ms), which is consistent with our expectation that edge devices are more resource-constrained than cloud devices.
As required by DisGB, each node has a copy of the DisGB registry containing \texttt{ip:port} tuples and broker areas.
All \gf{} Bridge instances use the local broker areas as their service areas, only the cloud-deployed instance subscribes to the whole world.

\subsubsection{\textbf{Distance/Latency Change Scenario}}
\label{subsubsec:distance}
\begin{figure}
    \centering
    \includegraphics[width=\linewidth]{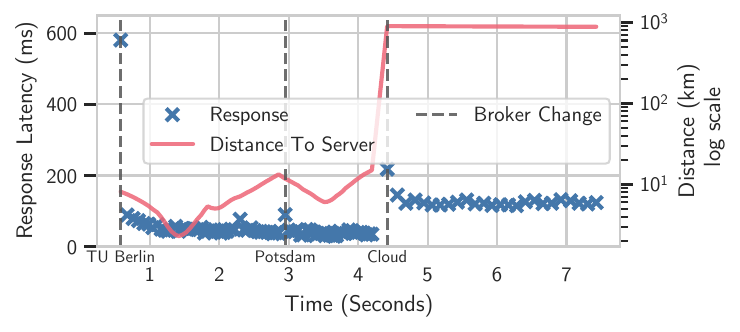}
    \caption{Each client's movement is followed by a function call, with response latencies plotted in blue (ms). The red line shows client-to-server distance (km), and vertical lines indicate switching to a new \gf{} node based on location updates. The client's movement is illustrated in \cref{img:exp-map}.}
    \label{img:exp-distance}
\end{figure}
Motivated by the goal of demonstrating geographical distribution transparency for clients, the Distance/Latency Change scenario evaluates the efficacy of geographical routing by analyzing the impact of client proximity to \gf{} node on latency, including handovers.
In this experiment we use both of the edge nodes, one located in Potsdam and one at TU Berlin.

A client device initiates function calls while traversing a path from northern Berlin to western Potsdam (see \cref{img:exp-map}).
Initially, the client connects to the TU Berlin edge node, and for the remaining 98 locations, it transmits a location update followed by a function call request.

\subsubsection*{Results}
As \cref{img:exp-distance} shows, there are peaks for the so-called cold start, switching to the Potsdam edge, and switching to the cloud, respectively.
The red line shows the distance of the client from the responding server, and vertical lines show  handover to another broker.
\cref{img:exp-distance} presents response times that encompass location updates.
While edge servers demonstrate consistent latencies, cloud responses exhibit expected increases due to physical distance.
This confirms the direct impact of server distance on response latency and the effectiveness of \gf{} in routing client requests to the closest FaaS server.

\subsubsection{\textbf{High Load Scenario}}
\label{subsubsec:highload}
\begin{figure}
    \centering
    \includegraphics[width=\linewidth]{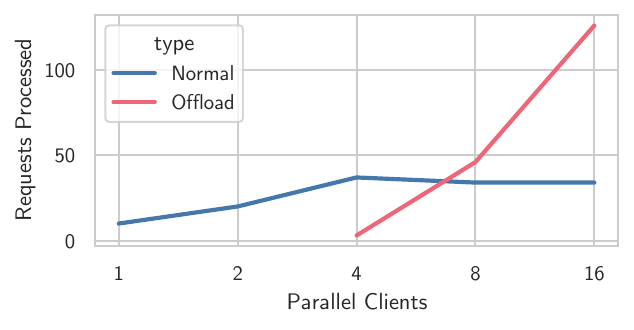}
    \caption{Breakdown of processed requests: Edge responses ("Normal") and cloud responses ("Offload") across various number of parallel clients. %
    }
    \label{img:exp-highload}
\end{figure}
This section demonstrates the ability of \gf{} to maintain continuous service transparently, ensuring clients do not need to intervene when requests are offloaded by \gf{}.
As the focus of this scenario is testing the offloading functionality, this experiment comprises just one edge and one cloud node.
To this end, we employ a computationally intensive function to push the edge server to its limits, thereby increasing the likelihood of requests being offloaded to the cloud for processing.
The function difficulty is increased by a factor of 100 to calculate prime numbers under 1 million on both servers to further overloading the edge server.
We also set the clients' ``result timeouts'' to be high to prevent them from retrying while their request is being processed.

In this scenario each non-moving client calls at a frequency of one call per second and makes a total of 10 calls.
The number of clients varies from 1, 2, 4, 8, to 16 for each experiment.
Each load setup involves 3 identical experimental runs, and as the outcomes were remarkably similar, we present just one result per load.
The clients' initialization is established using synchronized threads.
Each client call triggers launch of a new thread, independent of the completion status of the preceding request.
Moreover, \gf{} edge offloads requests either when receiving a bad HTTP response code (outside the 200 range), or the HTTP call has a timeout (10 seconds).
We measure how increasing the clients load affect the number of offloads.

\subsubsection*{Results}
\cref{img:exp-highload} shows the number of requests processed by the cloud (``Offload'') and by the edge (``Normal'') depending on the number of parallel clients (i.e., load).
As \cref{img:exp-highload} suggests, there are no offloads with 1 and 2 clients.
Offloading starts to occur in some 4-clients experiments.
We observe that for the 1, 2, and 4 parallel clients, the experiments take almost 10 seconds, meaning that each call is responded below or equal to 1 second.
For 8 and 16 clients, however, the experiments take about 18 and 25 seconds, respectively.
This experiment demonstrates \gf{}'s successful facilitation of transparent offloading for incoming requests via DisGB from edge nodes, guaranteeing uninterrupted client response delivery under load.

\subsubsection{\textbf{Outage Scenario}}
\label{subsubsec:outage}
\begin{figure}
    \centering
    \includegraphics[width=\linewidth]{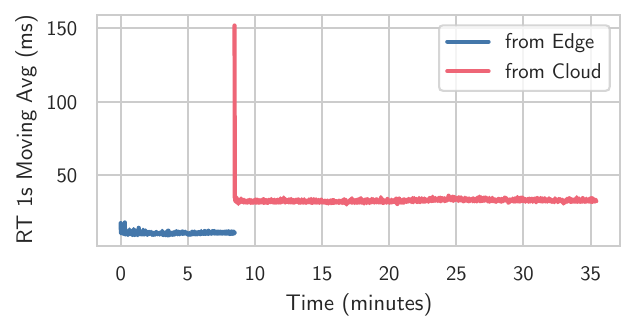}
    \caption{One-second moving average of response latency for 100k client calls: After 50k responses, edge node failure routes remaining requests to the cloud. Client timeouts during edge shutdowns necessitate retries.
    }
    \label{img:exp-outage}
\end{figure}
We chose this scenario to demonstrate that even in the event of a complete failure in an edge \gf{} Bridge while serving its clients, the system continues to operate, ensuring that clients receive a response from the cloud.

Similar to \cref{subsubsec:highload}, one single edge and one cloud node are used.
Since certain requests might reach the edge during its shutdown and remain unprocessed, \gf{} clients are allowed to retry once.
And the edge \gf{} is set to process 50,000 requests (half) and shutdown.

In this scenario, a stationary client initiates 100,000 call requests to the edge node. The edge node processes the first half of these requests before shutting down, after which the remaining requests are handled by the cloud.
Unlike the approach described in \cref{subsubsec:highload}, the client utilizes a wait-for-response strategy.

\subsubsection*{Results}
In \cref{img:exp-outage}, the outage is visible as a latency peak, indicating client retries when the edge Bridge receives a request during shutdown.
The blue and red lines depict one-second rolling moving average response times.
The edge (blue) exhibits a mean response time of 9 ms, while the cloud (red) demonstrates a mean response time of 29 ms.
Each experiment lasted approximately 35 minutes.
This experiment shows that a reliable DisGB can transparently route the requests to the cloud in case of a \gf{} Bridge failure.
Consequently, it successfully demonstrates that the \gf{} leveraged by a reliable DisGB, continues routing requests to the cloud, even in case of a \gf{} Bridge failure at the edge.

%% file: sections/5_conclusion.tex
\section{Conclusion and Future Work}
\label{sec:conclusion}
This work addressed the growing demand for geographically distributed computing by proposing \gf{}, a novel edge-to-cloud FaaS platform that prioritizes abstraction.
\gf{} leverages real-time client location information to transparently route function calls to the nearest available FaaS resource, minimizing network hops and latency.
By abstracting away detailed handover mechanisms, \gf{} provides a transparent execution experience, even with limited edge resources.
We developed and evaluated an open-source proof-of-concept prototype, demonstrating the feasibility and effectiveness of \gf{}.

In future work, we aim to optimize function placement to ensure functions are deployed in locations that minimize latency and maximize resource utilization.
Additionally, we will explore and experiment with optimized policies for FaaS server selection and cloud offloading based on factors such as utilization and resource availability.